\documentclass[Afour,sageh,times]{sagearxiv}

\usepackage{moreverb,url}

\usepackage[colorlinks,bookmarksopen,bookmarksnumbered,citecolor=red,urlcolor=red]{hyperref}

\newcommand\BibTeX{{\rmfamily B\kern-.05em \textsc{i\kern-.025em b}\kern-.08em
T\kern-.1667em\lower.7ex\hbox{E}\kern-.125emX}}

\usepackage{color,url, graphicx, multicol}
\newcommand{\furl}[1]{\footnote{\scriptsize \url{#1} \normalsize}}
\newcommand{\ie}{\emph{i.e.~}}
\newcommand{\eg}{\emph{e.g.~}}
\newcommand{\wrt}{\emph{w.r.t.~}}

\begin{document}

\runninghead{Adam and Arduin}

\title{An agent-based epidemics simulation to compare and explain screening and vaccination prioritisation strategies}

\author{Carole Adam (Univ. Grenoble-Alpes, LIG, Grenoble, France) \\ and Hélène Arduin (UMR IDEES, CNRS, Rouen, France)}



\email{carole.adam@imag.fr}

\begin{abstract}
This paper describes an agent-based model of epidemics dynamics. This model is willingly simplified, as its goal is not to predict the evolution of the epidemics, but to explain the underlying mechanisms in an interactive way. This model allows to compare screening prioritisation strategies, as well as vaccination priority strategies, on a virtual population. The model is implemented in Netlogo in different simulators, published online to let people experiment with them. This paper reports on the model design, implementation, and experimentations. In particular we have compared screening strategies to evaluate the epidemics vs control it by quarantining infectious people; and we have compared vaccinating older people with more risk factors, vs younger people with more social contacts. 
\\\textbf{Note:} This paper is an extended version of a conference paper presented at ISCRAM 2022 [1].
\end{abstract}


\maketitle

\section{Introduction}


The COVID-19 epidemics has now been lasting for over 2 years since the first cases in late 2019. The only control strategy for the first year was general lockdowns, but was hard to maintain longterm for economical and mental health reasons. In a second phase, when tests became available, the new strategy was to start large screening campaigns and to isolate (detected) sick individuals. Vaccines have then become available in December 2020, and have allowed to lift most constraining sanitary measures. 


However, with the epidemics lasting longer than expected and its end being hard to predict, people are tiring out, sanitary measures are not always well accepted, trust goes down \cite{strandberg2020coronavirus}. Fake news circulate with deadly consequences \cite{nieves2021infodemic}, such as refusing or hesitating to get the vaccine \cite{daly2021public, jennings2021lack}. We believe that it is very important to inform the population and explain the mechanisms of the epidemics and the reasons behind all measures \cite{home2020transparency}. Indeed, understanding constraints will improve their acceptability. 

We claim that an interactive simulator is a good tool to explain mechanisms by letting users play a role and learn by exploring what-if scenarios.
We have therefore designed an agent-based model 
that allows to \textbf{simulate} various screening and vaccination strategies on a virtual population, and to \textbf{compare} these strategies in order to discover insight about their optimal parameters. 
It is simple and interactive, and users from the general population can play with it in order to \textbf{understand} the complex mechanisms behind the epidemics and the reasons for the various sanitary measures. This work is part of a larger initiative aiming at answering questions from the general public about the COVID-19 epidemics, through interactive simulators along with explaining texts written by researchers from various disciplines \cite{covp-rofasss}.


\section{Challenges of screening} \label{sec:screening}

This section introduces some useful background about screening, quality features of tests, 
and possible prioritisation strategies for allocating a limited number of tests. 

\subsection{Lockdowns and screening}

When lifting the lockdown after the first COVID-19 wave in spring 2020, the main goals of many countries around the world were to get back to a less restricted way of living while still maintaining the epidemic under control to avoid a "second wave". Indeed, due to the low circulation of the virus during lockdown, herd immunity was still insufficient to prevent a rebound and new wave. For instance, it was estimated\footnote{by Pasteur Institute: https://www.pasteur.fr/fr/espace-presse/documents-presse/modelisation-indique-qu-entre-3-7-francais-ont-ete-infectes} that only 3 to 7 \% of the French population had been exposed to the virus (and was therefore immunised) when exiting the first French lockdown in May 2020. And as time has since proved, not only a second wave, but several more epidemics waves appeared, pushing some countries to enforce other lockdowns. 

But the general lockdown is hard to maintain on the longterm for both psychological and economic reasons \cite{atalan2020lockdown}, and is hard to lift without creating a new wave since herd immunity does not develop during lockdown. One solution is to selectively isolate only sick individuals. However, this strategy is hard to implement efficiently. Indeed, the COVID-19 incubation time is long (a week on average, but sometimes up to 20 days), so infected people have time to infect their contacts before they are detected and quarantined. Besides, the share of asymptomatic cases was still mostly unknown, but estimated to be around 30\% \cite{treibel2020covid}, meaning many infected people could unknowingly spread the virus among their contacts.

This implies that governments could not rely entirely on symptomatic displays to isolate infected people, but needed to \textbf{test} their population broadly by running large scale \textbf{screening campaigns}. This is precisely the strategy recommended by the World Health Organisation (WHO), as early as the 16 March 2020\furl{https://www.who.int/dg/speeches/detail/who-director-general-s-opening-remarks-at-the-media-briefing-on-covid-19---16-march-2020}: to test any suspicious case to confirm potentially infected individuals; to trace their contacts in order to identify chains of contamination; and isolate only (potentially) infectious people. But it took time to develop reliable tests and start this strategy.


\subsection{Quality of tests}

There exists different types of tests to detect the SARS-COV-2 virus responsible for COVID-19, in particular PCR (polymerase chain reaction) tests, serological tests, antigenic tests, and auto-tests that one can realise at home. These tests have different levels of quality, depending on 2 factors:
\begin{itemize}
    \item \textbf{Sensitivity} of a test indicates the probability that the test is positive when the tested person is really sick. A 100\% sensitive test applied to a sick individual will always return positive; therefore a negative test gives absolute certainty that the tested individual is indeed not sick. In other words, there are no false negatives with a 100\% sensitive test; so sensitivity is the proportion of true negatives. 
    \item \textbf{Specificity} of a test indicates the probability that the test is negative when the tested person is really not sick. A 100\% specific test applied to a non sick individual will always return negative; therefore a positive test gives absolute certainty that the tested individual is indeed sick. In other words, there are no false positives with a 100\% specific test, so specificity is the proportion of true positives.
\end{itemize}

However, it is impossible to design tests that are perfect on both criteria (or even on a single one). Screening tests always have an error margin. In particular, screening tests cannot be both highly specific and highly sensitive, so a compromise must be found between two opposites:
\begin{itemize}
    \item Very \textbf{sensitive} tests are more likely to be positive with sick individuals: this reduces the rate of false negatives, so prevents missing infected people who keep moving around instead of being quarantined;
    \item Very \textbf{specific} tests are less likely to be positive when the individual is not sick: this reduces the rate of false positives, to prevent from quarantining healthy people.
\end{itemize}

The first screening tests designed for COVID-19 were relatively specific (in the range of 95 to 98\% of true positives) but still little sensitive (sometimes up to 30 to 40\% of false negatives, sick but not detected by the test). As a result, it was sometimes necessary to do a second test to confirm a negative test result.

\subsection{Screening objectives}

Time was needed to develop reliable quality tests and increase testing capacity. As a result, testing kits were rare at the start of the epidemics, forcing governments to prioritise who should be tested first to optimise the impact of the testing campaign. Even nowadays, when testing kits are widely available, and as new variants of the virus circulate very fast, the number of daily tests has exploded, posing a new issue of financing those tests. Some countries therefore again choose to restrict tests to some categories of people, for instance, the elderly who are more at risk of serious forms, or people with symptoms. Other countries require non vaccinated people to pay for the tests, also in order to limit the number of tests performed. 

Screening tests actually pursue two main (partly contradictory) goals.
\begin{itemize}
    \item The first one is to \textbf{control} the epidemics, by spotting infected people and isolating them to break contamination chains. 
    \item The second one is to \textbf{know} the epidemic, \ie evaluate as precisely as possible the total number of people infected at a given time, and deduce the actual case fatality rate. 
\end{itemize}
These goals involve different screening strategies: in order to best control the epidemics, one should test in priority people who are more likely to carry the virus, but this leads to an over-estimation of the global circulation; to best know the epidemics, one should randomly test a representative sample of the global population, but this will lead to a large number of negative tests, failing to isolate many infected people. The best screening strategy is therefore not intuitive.

\subsection{Screening prioritisation strategies}

Under the constraint that testing kits are in limited supply, governments want to prioritise wisely who should be tested, in order to reach both goals with the minimum amount of tests. For instance, France started testing late and slowly\furl{https://www.usinenouvelle.com/article/en-retard-la-france-monte-en-puissance-pour-les-tests-de-diagnostic-du-covid-19.N945261}: it took some time to design reliable tests, and the small number of such available tests was thus limited to healthcare workers and people at risk. Nowadays, tests are widely available and are the most cost-effective mitigating measure \cite{rezapour2021economic}, but some countries start restricting them again in order to limit the financial cost for society, for instance, by reserving them to elderly people, or by asking non-vaccinated individuals to pay for the tests.

The various possible targeting strategies have different impacts on both goals stated above:
\begin{itemize}
    \item \textbf{Random targeting} consists in choosing randomly people who should be tested. This is a more representative sample of the population, and provides better knowledge of the current state of the epidemics. But when the incidence of the virus is very small (as it was after the first lockdown), the proportion of people infected is very low, so most tests will return negative. There is therefore a risk of "wasting" many tests, \ie the chances of finding infected people to isolate them and control the epidemics are low.
    \item A solution is to \textbf{target suspicious cases} (the symptomatic ones), but this strategy is insufficient to control the epidemics since it ignores all the (also contagious) asymptomatic cases. Besides, the sample is not representative of the general population, and the high proportion of positive tests in the sample might lead to overestimate the global circulation of the virus.
    \item Another strategy consists in \textbf{targeting people who work outside of home}, since they are more likely to get infected and/or infect others. For instance, at the beginning of the epidemic, healthcare workers were tested in priority, since they were the most exposed to the virus; in order to reopen schools, there was also a focus on testing teachers, school workers, and now all the children from the same class as an infected pupil. This strategy focuses on controlling the epidemics while allowing for economic activity, but it ignores contaminations that happen outside of work (shopping, leisure...).
    \item Finally a last interesting strategy consists in \textbf{targeting high-risk people}. Their profile is now better known, in particular elderly people or people with comorbidities \cite{jordan2020covid}. The goal of this strategy is to detect infection soon and treat them early to prevent serious complications. But the results would then not be representative of the global circulation of the epidemic in the general population.
\end{itemize}

The choice is not intuitive, and we claim that simulation can help compare different strategies in order to draw interesting insight. Indeed, simulation allows to run the exact same scenario with only the parameters of the screening campaign varying, which is impossible in reality, and to compare estimations with the ''real'' epidemic curve, which can be known only in a simulation.

\section{Challenges of vaccination} \label{sec:vaccine}

Vaccines became available in December 2020, and many different ones are now available. This section provides some useful background related to our model.

\subsection{Vaccination}

The goal of vaccination is to provide people with some level of immunity against the virus, that will protect them from infection, and ultimately to create collective immunity at the level of the population to stop further propagation of the virus. The impact of the vaccination campaign can be measured on different indicators: the height of the epidemic peak (how many people are sick at the same time at the peak), the duration of the epidemics wave (how long it takes before nobody is sick anymore), the total number of people who got sick, the total number of serious cases, the number of people in hospitals, or the number of deaths. The efficiency of a vaccine can be defined on different terms \cite{hodgson2021defines}, such as reducing the risk of infection, the severity of the disease, or the duration of infectivity. 
A vaccine can have different levels of efficiency on these different aspects. It was initially hard to know exactly which factor was impacted by the vaccine: its effect on infection was tested by trials before release, but it was not clear if it was also effective on transmission \cite{lipsitch2021interpreting}, on asymptomatic forms \cite{bleier2021covid} or on the risk of serious forms. Governments sometimes made simplified statements to encourage people to get the vaccine\furl{https://www.liberation.fr/checknews/transmission-du-covid-19-les-autorites-ont-elles-menti-sur-lefficacite-du-vaccin-pour-justifier-les-pass-sanitaire-et-vaccinal-20221014_JEAR5KU73RFNTPRDWQCLHSNZQE/}. The efficiency of the vaccine has also decreased against the new variants of the virus \cite{bian2021impact}. 


\subsection{Individual vulnerability}

Literature has shown that infections and serious forms are more frequent in people with risk factors, in particular those aged above 60 \cite{romero2020age} and/or having comorbidities (often associated with age) such as diabetes or chronic illnesses \cite{ejaz2020covid, bajgain2021prevalence, elezkurtaj2021causes}.

Literature also shows that younger people have more contacts in average than older people. For instance \cite{ibuka2016social} studied influenza in Japan and showed difference in contact patterns based on age, gender, as well as during the week vs holiday. \cite{backer2020social} also study the role of contacts with people of different age groups; they conclude that the case growth rate increases when there is more contacts with elderly people, but decreases with contacts among the same age group.

\subsection{Vaccine prioritisation strategies}

The first vaccine against COVID-19 was released in December 2020. Developed countries invested a lot of money to secure enough doses for their population, but the production and injection of millions of doses takes time. Furthermore, the immunity conferred by the vaccine only lasts for a few months \cite{dolgin2021covid}, so it is necessary to inject boosters regularly. As a result, governments are faced with a choice about who should be vaccinated first:
%
\begin{itemize}
    \item Vaccinating in priority people with \textbf{comorbidities} (elderly people or people with other health factors increasing their vulnerability), in order to protect them from serious forms of the illness;
    \item Vaccinating in priority people with \textbf{more contacts} (such as health workers who are in contact with many sick or vulnerable people, teachers, or younger people who have more social contacts), in order to reduce the global transmission of the virus.
    \item Randomly vaccinating the entire population, without any order of priority.
\end{itemize}
Similar to screening, the optimal strategy is not intuitive since different strategies improve different indicators. One might reduce the number of serious forms but leave many younger people exposed, or reduce the total number of contaminations to prevent saturation of the healthcare system but take the risk of more serious forms among vulnerable people. Therefore, once again simulation could help in testing strategies by varying their parameters in isolation.

\section{Agent-based simulation of epidemics} \label{sec:soa}

\subsection{Goal of this work}
One can see from the background above that finding the best (screening or vaccination) strategy on all accounts is not easy by using only intuition. We claim that simulation can help by allowing us to compare different strategies and to measure their effects. In particular, simulation will allow us to run the exact same scenario several times, with only selected parameters (prioritisation strategy, etc) varying, which is impossible in reality. 
In this section we discuss existing simulations of the COVID-19 epidemics, mitigating interventions in general, and screening and vaccination in particular.
The main measures against COVID-19 have been quarantine, contact tracing, screening, and isolation; there is no consensus on best practices, and countries differ in their approach, but a survey of medical publications \cite{girum2020} shows their efficiency, in particular when combined together. It is hard, however, to compare their efficiency in real life since their individual impact cannot be isolated, but simulation can help comparing them ''in silico''.

\subsection{Advantages of simulation}
Computer simulation has many benefits in the context of the crisis management in general, and the current epidemics in particular. Indeed, only simulation allows to compare different intervention strategies, all other things being equal. In the real world, we can only compare between different countries applying different strategies, but they also differ on other regards: climate, culture, etc, that all might influence virus spread, so it is impossible to isolate the precise impact of the strategy being evaluated. But in the simulated world, we control all parameters, and repeat the exact same experiment with only the strategy changing, in order to evaluate its impact independently of all other factors.

Besides, in this work, we are interested in evaluating screening strategies and their efficiency to assess and control virus spread. In the real world, there is no way to access the actual number of infected people (unless we could simultaneously test the entire country), so it is impossible to evaluate how good the curve estimated from the tests is compared to the "real" curve. On the contrary, in a simulation we know the epidemiological status of all agents, so we can access the \emph{real} (simulated) epidemic curve, and compare it with estimations obtained from various screening strategies. Simulation is therefore a great tool to assess screening strategies on a virtual population.

\subsection{Approaches to epidemics simulation}
As a result, a lot of models of the COVID-19 epidemics have been published in the last 2 years. Current epidemic models mainly fall in two approaches. 

\textbf{Compartmental models} divide the population in a number of epidemiological classes. The simplest ones (often shortened as SIR models) use only 3 compartments: Susceptible (not yet infected so not immune), Infected (and contagious), Recovered (and immune). The hypothesis is that recovered individuals are then immune and cannot get infected again, which has proven wrong for COVID-19. Compartmental models of COVID-19 have often integrated more compartments, such as Asymptomatic or Hospitalised, in order to more precisely represent the dynamics. These models then rely on the mathematical resolution of differential equations to give a \textbf{macroscopic} view of the epidemic dynamics. It is therefore quite fast and scalable.

\textbf{Agent-based models} model each individual as an autonomous agent, in order to give a \textbf{microscopic} view of the situation. Agents are heterogeneous, initialised with different values of their attributes, such as age, gender, comorbidities, or social behaviour. This allows to model the influence of individual decisions on the virus spread, such as a refusal to get vaccinated, or not respecting social distancing. These models are more complex to initialise as they require behaviour data that is hard to get, and are less scalable since they require to compute the behaviour of each individual agent. However, they are more precise, in particular to study why a specific individual got infected.

Both approaches have their benefits and drawbacks depending on the goal and scale of the simulator. Our goal being to explain the complex mechanisms behind the epidemics rather than predicting its spread at the scale of the country, we chose an agent-based approach. We survey various existing agent-based models in the next paragraphs.

\subsection{Simulations of interventions}

A very early report (March 2020) about the results of an epidemiological modelling informed the first policy-making in the UK and other countries \cite{ferguson2020report}, leading to general lockdown. At that time, in the absence of a vaccine, they assessed other non-pharmaceutical interventions to reduce contacts in the population and therefore virus transmission, with 2 main possible strategies: mitigation aiming only at \textbf{slowing down} the epidemic spread, vs \textbf{suppression} aiming at completely stopping it. Their results showed that optimal mitigation policies might reduce the pressure on healthcare system by two thirds and deaths by half, but might still result in hundreds of thousands of deaths and overwhelmed healthcare systems. Therefore they suggest that where possible, countries should aim at suppressing the epidemics rather than just mitigating it, which would require a combination of several very constraining measures (quarantines, distancing, school closures) despite their negative side effects. Besides, they predict that such measures would need to be maintained indefinitely (or until a vaccine is found) to prevent a rebound as soon as they are relaxed. Not knowing if such suppression can be maintained on the long-term, or how to reduce their social and economic costs, they also suggest intermittent measures: closely monitoring the epidemic progression to temporarily relax measures, but re-instantiate them when needed. This is similar to what happened around the world in the past 2 years.

Other countries have also developed simulators to help policymakers. For instance, COMOKIT \cite{gaudou2020comokit} is a framework composed of several realistic spatialised agent-based models of the epidemics and of various interventions, aiming at informing public health decisions made by the Vietnamese government. These models have also been applied to towns in other countries, such as Nice in France \cite{chapuis2021using}. The models run on the GAMA platform. They can be fed from various data sources, provided by the Vietnamese government (census data, epidemiological data) or by private actors (Facebook data, mobile phone data). They have been used to quickly compare potential strategies (lockdown at different scales, quarantines), as well as to suggest optimal timing or combination of multiple strategies. The models offer a precise representation of the population and of contamination in closed spaces (shops, schools) although they neglect the role of transportation. They are very complex models aimed at guiding policymakers, rather than informing the general public.

Covasim \cite{kerr2021covasim} is an agent-based model of the epidemics that can be tailored to various local contexts (\eg age distribution, daily contacts, epidemic progression in number of reported cases and deaths) and has been actually used in a number of countries. It allows to test several types of interventions: physical (\eg lockdown), diagnostic (\eg screening or contact tracing), or pharmaceutical (\eg vaccination). However, it is targeted at researchers and policy makers rather than the general public, and as such is much more complex than our intended simulator. Moreover, testing strategies are expressed in terms of probabilities of testing people with or without symptoms, in/out of quarantine, or over a certain age; it does not include other interesting strategies such as random sampling or prioritising essential workers.

\subsection{Simulations of testing}

Various models have specifically studied different screening strategies. For instance, \cite{paltiel2020assessment} modelled several scenarios regarding the measures needed for a safe re-opening of US colleges. They modelled 5000 students, of whom 10 were infected at the start of the semester, and tested several epidemic scenarios (with reproductive numbers values between 1.5 and 3.5, a 0.05\% fatality rate, and a 30\% probability of showing symptoms when infected). They varied the following parameters of screening: frequency (every 1, 2, 3, or 7 days), sensitivity of tests (between 70 and 99\%), specificity of tests (98 to 99.7\%), and cost (10 to 50 dollars per test). They conclude that it is best to screen frequently (every 2 days) in addition to strict observance of sanitary measures to keep the reproduction number under 2.5. In order to limit costs, they also show that even a rapid, less expensive but poorly sensitive test (around 70\%) is sufficient to control the number of infected students, who are isolated in a dedicated dormitory (within an 8h delay). 
Of course this strategy is not applicable at the scale of a country (not enough medical staff to test the entire population every 2 days). The population of a college campus is also not representative of the general population, and in particular is younger so less exposed to serious forms.

\cite{atkeson2020economic} focus on the economic benefits of testing. They provide an SIR model of the US population with 5 age groups, working in various economic sectors. Their study shows that the economic benefits of rapid screening programs far exceed their costs, with a ratio between 4 and 15 (excluding the monetised value of lives saved) depending on the parameters of the screening. However, an interesting aspect of their study is that they consider a variable adherence to quarantine measures, depending on the probability to be a false positive: 
They conclude that tests used must be highly specific, or combined with a confirmatory test, in order to both reduce the cost of having healthy workers in quarantine, and decrease the number of people who break quarantine because they (wrongly) believe they are a false positive. 

To conclude, models that include testing as an intervention generally focus on \textbf{controlling} the epidemics: testing is part of the testing, tracing, isolating strategy recommended by WHO. On the contrary, we want to compare testing strategies with respect to 2 different objectives: not only controlling the epidemics by appropriately isolating infected individuals, but also precisely estimating the current state of the epidemics (number of cases), which is useful to evaluate the impact of current measures and the necessity to adapt them.

\subsection{Simulations of vaccination}
Finally, other models focus on the impact of a vaccination campaign. This is relevant as vaccination poses a similar problem of prioritisation for the allocation of a limited number of doses, but also an additional problem of compliance, or trust: some people might not want to get vaccinated, for various reasons \cite{hornsey2018psychological}. 

For instance, \cite{li2021returning} modelled a nation-wide vaccination campaign in the USA. They varied parameters such as vaccine efficacy or population compliance, and tested 6 scenarios combining a vaccination campaign with other interventions (distancing, etc). Their results show that the vaccine significantly reduces infections, even with very low population compliance. However, they also show interesting counter-intuitive results: when compliance is very high, and since the older population is vaccinated first, the delay is longer for younger and more connected individuals to access the vaccine; as a result, the virus spreads more than if the compliance was lower. This also proves that the vaccine alone is not sufficient, and should be combined with other interventions to reduce the spread of the epidemics. 

Other studies are concerned with prioritising the vaccines. \cite{priovax21} simulate different prioritisation strategies with a limited supply of vaccines, in a US urban region of 2.8 million residents. They show a limited impact of vaccination on reducing viral spread, mainly due to exponential contaminations before the start of vaccination, meaning that lots of adults were already immunised. Consequently, they suggest that vaccines should be distributed as fast as possible among all eligible adults after vaccinating the most vulnerable, to improve the speed of the vaccination campaign, rather than strictly respecting a priority schedule. The study of \cite{yang2021assessing} compares age-related vaccination strategies in 3 countries; the authors find that vaccinating younger people first allows to reduce the number of infections, while vaccinating elderly people first reduces the number of deaths.

The results of both these studies are consistent with the findings of \cite{li2021returning}. Besides, these studies also show the interest of simulation to reveal unexpected and unintended consequences of potential sanitary policies, that cannot be tested in real life.

\subsection{Summary}
Not many models specifically focus on comparing various screening strategies to both evaluate and control an epidemic. This can be explained by the now wide availability of testing kits in developed countries, that allow to massively test the entire population. Nevertheless, such work is still important to explore screening strategies in countries where tests are not yet widely available, or to reduce the cost of screening, or for potential future epidemics. Such a model can also be extended to comparing the allocation of other limited resources, such as the vaccine doses. 

In this paper, we present a model for comparing strategies of priority for the allocation of testing kits and of vaccine doses. Our model does not aim at predicting the exact evolution of the epidemics for deciders, but rather at \textbf{explaining} the mechanisms to the general public. It does so by letting users interact with the simulated population and take on the role of public health deciders to select the parameters of a screening strategy or a vaccination campaign. The originality of our work is to target the general public and focus on explaining a complex phenomenon, unlike most models that are targeted at governments to help them make decisions.


\section{Our model for screening strategies} \label{sec:model}

As explained above, our agent-based model is quite simple since it is targeted at the general population, and its goal is \textbf{not to predict} the evolution of the epidemics, but to explain its mechanisms.

\subsection{Characteristics of the population}

We have modelled a population of 2000 individuals, distributed in several age categories. This influences their sensitivity to the virus (older people have more risk to be symptomatic) as well as their mobility. For instance, 50\% of people aged 20 to 65 have work outside their home ('essential' workers) whereas the rest stay at home (remote work, furloughed workers, family carers, etc.). People aged less than 20 and over 65 are considered homebound (by respectively remote schooling and retirement). Individuals in our population can be in one of five distinct states regarding the virus, as shown in Figure~\ref{fig:state} below:
\begin{itemize}
    \item \textbf{Susceptible}: they have never been infected and are therefore not immune either
    \item \textbf{In incubation}: they have been infected but are not yet sick (it lasts 6 days on average)
    \item \textbf{Asymptomatic}: they are sick but display no symptom. Only a test will reveal them (30\% of patients below 65 years old, for an average of 21 days)
    \item \textbf{Symptomatic}: they are sick and display symptoms. (70\% of patients below 65 years old, and 100\% of older patients, for an average of 21 days)
    \item \textbf{Recovered}: they are immune and cannot get infected again (this was a hypothesis of our model, at a time when this was unknown \cite{roy2020covid}; we now know that reinfection after recovery or vaccine has a lower risk but is still possible \cite{cavanaugh2021reduced}
\end{itemize}

\begin{figure}[hbt]
    \centering
    \includegraphics[scale=0.2]{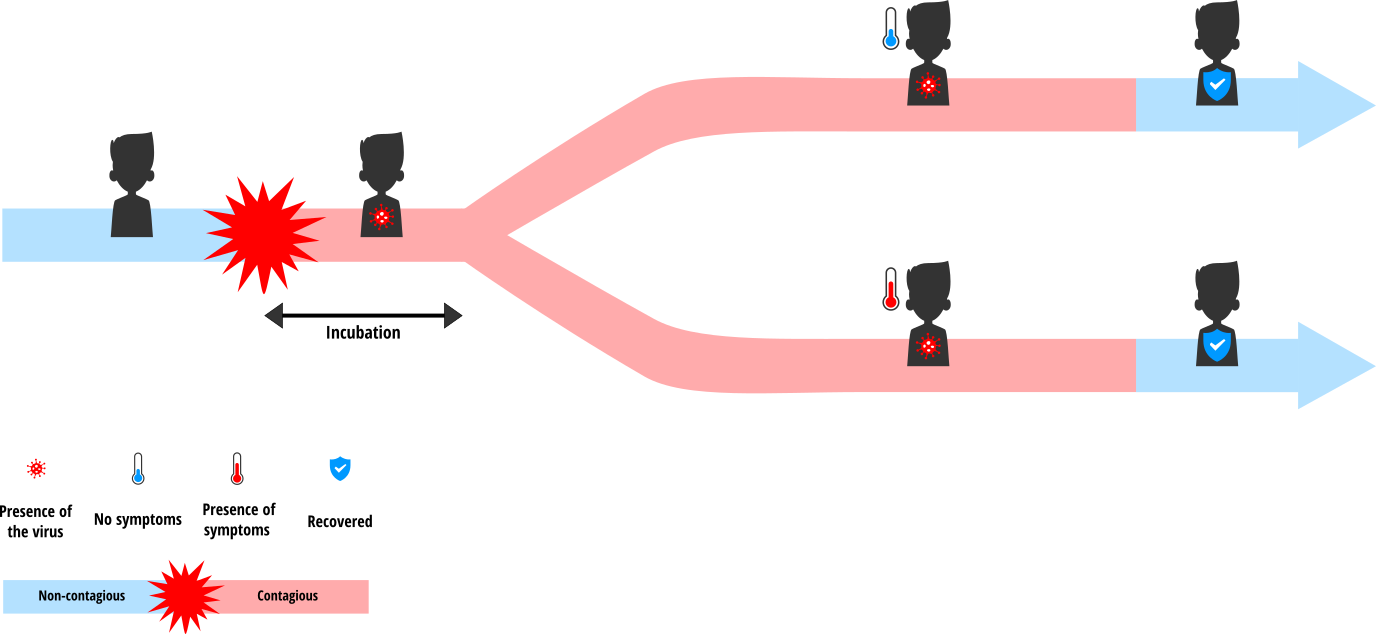}
    \caption{Evolution of epidemiological status of agents}
    \label{fig:state}
\end{figure}

When they move in the simulation world, individuals can come into contact with one another. Infected people (in incubation, asymptomatic and symptomatic) are all contagious and can therefore transmit the virus to susceptible people they are in contact with. Individuals working from home do not move but they can be in contact with people passing by their home (deliveries, postal services, etc.). However, they have less contacts on average than people who have to work outside of home.

In order to reduce the spread of the virus, people who tested positive are put into quarantine and are completely isolated: they cannot transmit the virus any further.

\subsection{Testing strategies}

In our model, we simulate tests with both a sensitivity and specificity of 90\%. It would be interesting to vary these 2 parameters in future work, to study their impact on the evaluation of the epidemic curve. Our goal is to find out how we can best use the tests available each day to reach the two main objectives of a massive testing campaign: to monitor the epidemic (optimise fit between estimated and ''real'' curve) and to control it (minimise the epidemic peak). 
The screening campaign has the following attributes:
\begin{itemize}
    \item The \textbf{number of tests} available each day. The testing strategy of the French government plans for 5 to 700.000 tests per week, which corresponds to about 2 to 3 tests per day for our 2000 agents;
    \item The \textbf{triggering time} for testing, defined in terms of a threshold X for the proportion of symptomatic people: when more than X\% of the population is symptomatic, the testing campaign starts;
    \item The \textbf{target population} being tested, among: random sampling; symptomatic people; elderly / at-risk people; or people working outside of home.
\end{itemize}

\subsection{Interactive simulator}

\begin{figure}[hbt]
    \centering
    \includegraphics[scale=0.2]{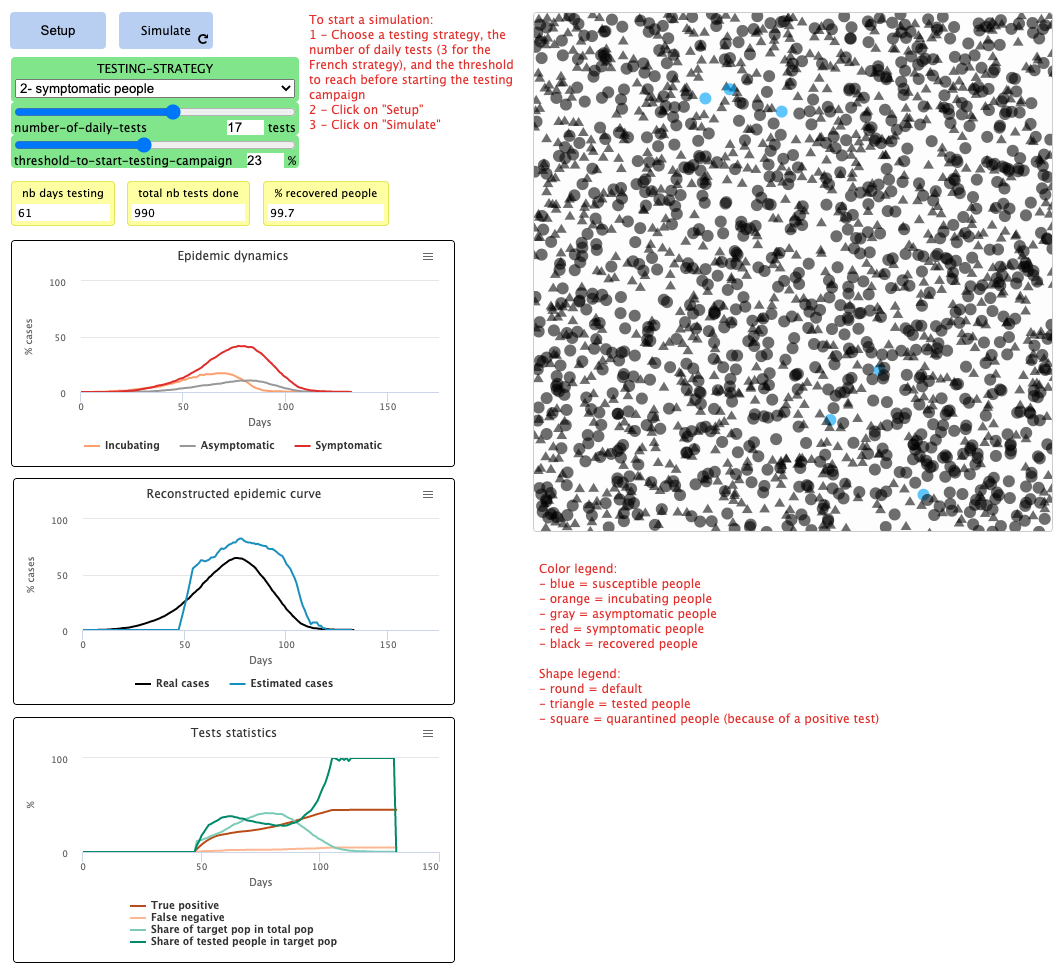}
    \caption{Interface of the online simulator}
    \label{fig:screen}
\end{figure}

Our simulator is implemented in Netlogo and playable online\footnote{Covprehension question 17: \url{https://covprehension.org/en/2020/05/12/q17.html}}. Fig~\ref{fig:screen} illustrates its online interface. The user plays the role of a public health decider with several goals:
\begin{itemize}
    \item To minimize the total number of tests used;
    \item To minimize the spread of the epidemic, \ie the total number of people infected;
    \item To minimize the number of people quarantined for no reason, \ie the number of false positive tests;
    \item To estimate as closely as possible the “true” curve in order to precisely monitor the epidemic over time;
\end{itemize}

\noindent To reach these goals, the user can modify the parameters of the testing campaign as defined above: \begin{itemize}
    \item The number of tests available each day (\ie time necessary to test the entire target population);
    \item The starting date of the testing campaign: immediately from the start of the epidemic, or later on;
    \item The target population: in-person workers, elderly, random, symptomatic. 
\end{itemize}

These goals are obviously not always compatible, and actions can have opposite effects on different goals, improving one at the cost of failing another. For instance, to best know the status of the epidemics, one could test every citizen but would fail to minimise the number of tests used; or one could randomly test people to get a representative picture of the situation, but would then ``waste'' many (negative) tests that fail to spot and isolate infected people. Managing the situation thus requires finding some compromise. We expect that playing with the simulator will help people understand the stakes behind the sanitary measures, and make them less subject to blindly believing disinformation. This will require future experiments of our simulator with users to test its actual impact. 

In the following section
we describe the experiments that we ran by manually varying the different parameters (target population, number of tests per day, and starting date), and we compare how well they estimate the ''real'' curve, which the simulation allows to know. 

\section{Screening experiments} \label{sec:expe}

\subsection{Inference of the epidemic curve}

There are several ways to compute the estimated curve from the results of screening tests. The number of confirmed cases communicated every day by the authorities of different countries is a combination of test results and expertise of health professionals. But like any statistical estimation, these figures have error margins and potential biases. We believe that the general public should be aware that the “true” number of infected people is unknown and can only be approximated. It is one goal of our simulator to allow them to observe these variations and errors.

Indeed in our simulator we know the status of all of the agents, so we do know this "true" number of infected agents over time, which is impossible in reality. We can therefore \textbf{compare the curve estimated using tests, with the “\emph{true}” curve}, in order to verify the correctness of the estimations obtained with different strategies. This is a great advantage of agent-based simulation, which will allow us to evaluate how well different computation methods do fit this real curve. We have actually tested two simplified estimation methods, described below: proportionality, and predictive values.

\subsubsection{Proportionality rule}
First, we could intuitively use a simple cross multiplication: the number of positive tests among the total number of tests provides an estimation of the proportion of cases in the general population, by a proportionality rule. For instance, if we test 700000 people and that 1000 of them are infected, this is a proportion of 0.14\%, so we deduce that out of the 70 millions residents of France, 100000 are infected. But this simple \textbf{proportionality rule} does not work well in epidemiology, especially when the number of people tested is low, or when the prevalence of the epidemic (\ie the total number of cases at a given time) is too low, as is the case at the beginning and at the end of an epidemic, or when tests are not entirely reliable.

\subsubsection{Predictive values}
Another method for estimating the total number of infected people is to compute the \textbf{predictive values} of the test, which depend on three elements: the prevalence of the epidemic, the rate of false positives (which is equal to $1-specificity$, with specificity being the probability that a positive individual receives a positive test), and the rate of false negatives (equal to $1-sensitivity$, since sensitivity is the probability that a negative individual receives a negative test) obtained with the test. The positive and negative predictive values are not intrinsic to the test (unlike true positive rate and true negative rate), and do not concern a single individual but a population, so they also depend on the prevalence of the epidemics in this population (\ie the probability that an individual is positive).

The positive predictive value (PPV) is also called precision. Its complement is the false discovery rate, \ie the rate of false positives among the total of positive tests. 
\begin{equation}
PPV = \frac{sensi * preval}{sensi*prev + (1-specif)*(1-preval)}
\end{equation}

The negative predictive value (rate of true negatives in total negative tests) is the complement of the false omission rate (rate of false negatives upon total negative tests).
\begin{equation}
NPV = \frac{specif * (1-preval)}{specif*(1-preval) + (1-sensi)*preval}
\end{equation}

The following paragraphs describe our experiments on a simulated population of 2000 agents, exploring different scenarios by varying the different parameters of the screening campaign.

\subsection{Comparison of estimated cases for the different samples of population}

In this scenario we choose a daily number of tests equivalent to that of the French government (\ie 3 tests for 2000 individuals) and start testing as soon as the first case appears (which has not been the case in France). 

\begin{figure*}[ht]
    \centering
\begin{minipage}[b]{0.45\linewidth}
    \includegraphics[scale=0.25]{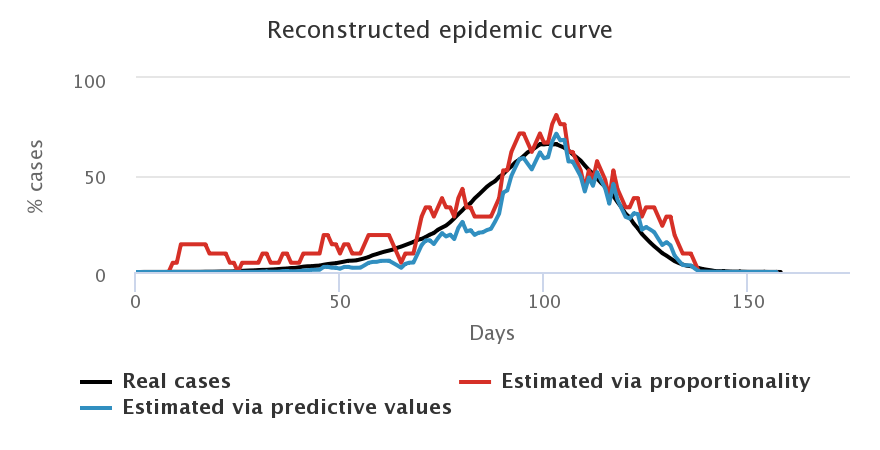}
    \centering \textbf{(a) Random sample}
\end{minipage}
 \hfill
\begin{minipage}[b]{0.45\linewidth}
    \includegraphics[scale=0.25]{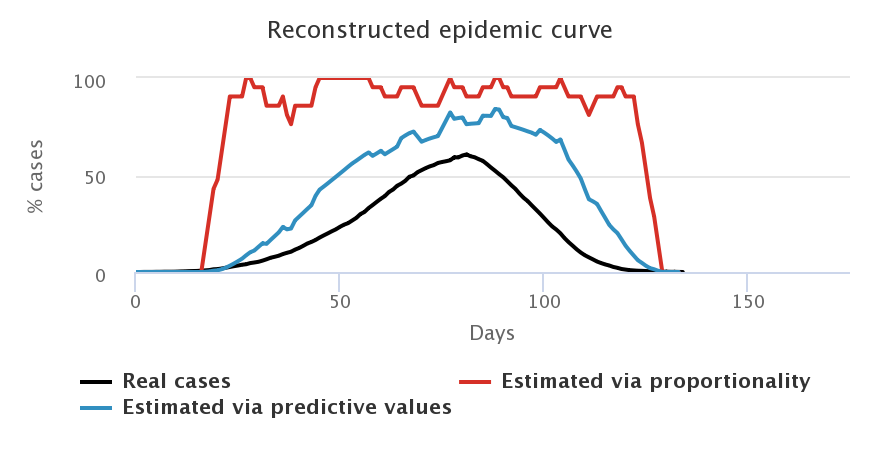}
    \centering \textbf{(b) Symptomatic sample}
\end{minipage}

\begin{minipage}[b]{0.45\linewidth}
    \includegraphics[scale=0.25]{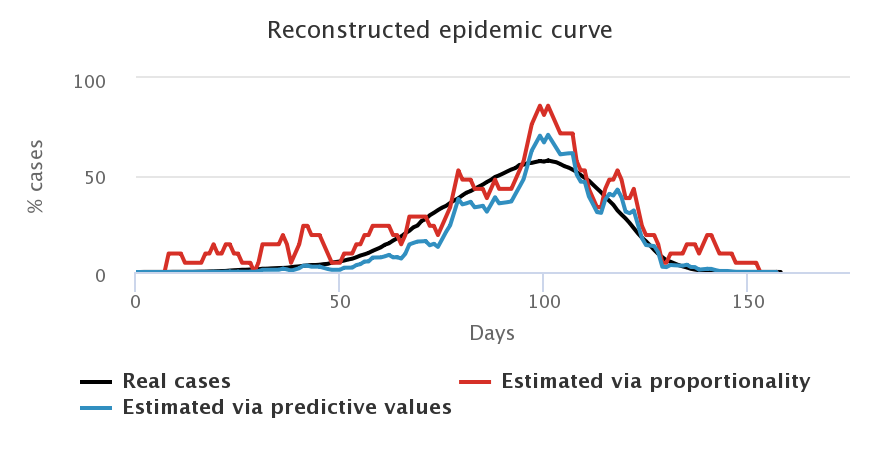}
    \centering \textbf{(c) Elderly sample}
\end{minipage}
\hfill
\begin{minipage}[b]{0.45\linewidth}
    \includegraphics[scale=0.25]{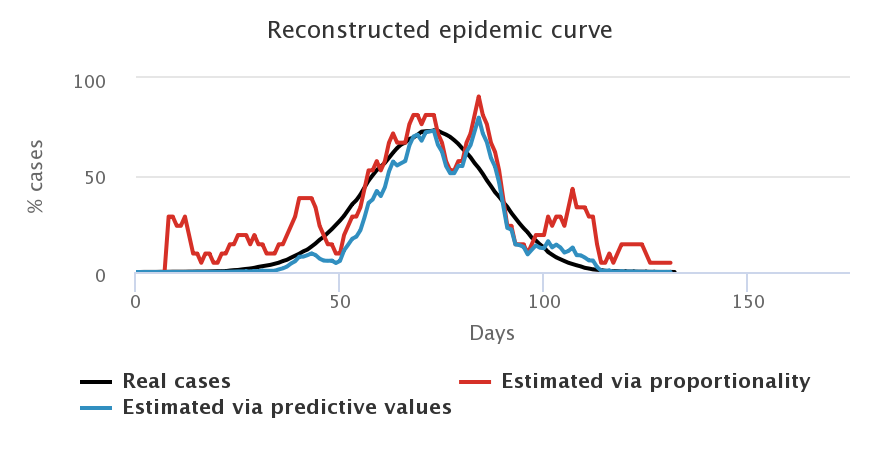}
    \centering \textbf{(d) Workers sample}
\end{minipage}
    \caption{Experiment 1: 3 tests/2000, immediate start, different samples}
    \label{fig:exp1}
\end{figure*}

We compare the curves representing the number of cases estimated by the proportionality rule (in red), the number of cases estimated by computing the predictive values (in blue), and the “true” number of cases (in black), for each sampling strategy (testing priority among: random, elderly / at-risk, workers, symptomatic). Figures~\ref{fig:exp1} summarises the results.

Firstly, we note that the blue and red curves (estimated cases) vary a lot more than the black curves (“true” cases). Indeed, they depend on the number of positive tests each day, which varies greatly depending on \textbf{who} is tested. In general, we choose to “smooth” this estimation by computing the mean result of tests over several days (here, a mean over 7 days), but there is always a larger variability depending on who is tested each week. For example, a lot of negative tests may result in an estimation of a decreasing epidemic which is not necessarily true (maybe we just tested non-infected people, which does not mean that no one is sick anymore).

We also note that the blue curves (with predictive values) are better estimators than the red curves (proportionality rule), especially at the beginning and at the end of the epidemic when the prevalence of the virus is low, and especially for symptomatic people, who are less representative of the total population.

Secondly, we find that the worst estimation happens when we only test symptomatic people (top right figure): since we only test symptomatic people, who therefore have a high probability of being infected by COVID-19, we obtain a high positivity rate of the tests, so we overestimate the “true” number of cases in the general population. Given that the tests we model are diagnostic ones, it is really not sound to extrapolate the number of cases in the general population from the tests performed on symptomatic people. The model confirms this. Such tests allow to control the epidemic by confirming and isolating infected people, but they do not allow to monitor its spread in the population.

\subsection{Comparison of estimated cases \wrt the number of daily available tests}

\begin{figure*}[hbt]
    \centering
    \begin{minipage}[b]{0.3\linewidth}
        \includegraphics[scale=0.22]{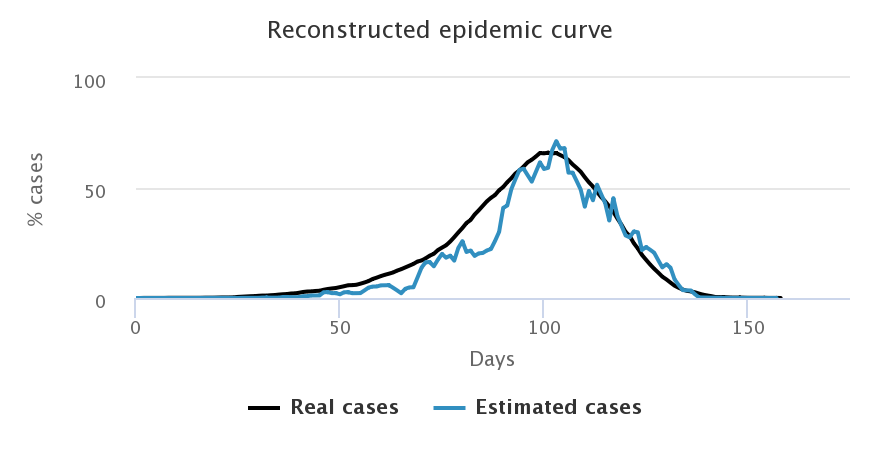}
        \centering \textbf{(a) 3 tests / 2000 people}
    \end{minipage}
    \begin{minipage}[b]{0.3\linewidth}
        \includegraphics[scale=0.22]{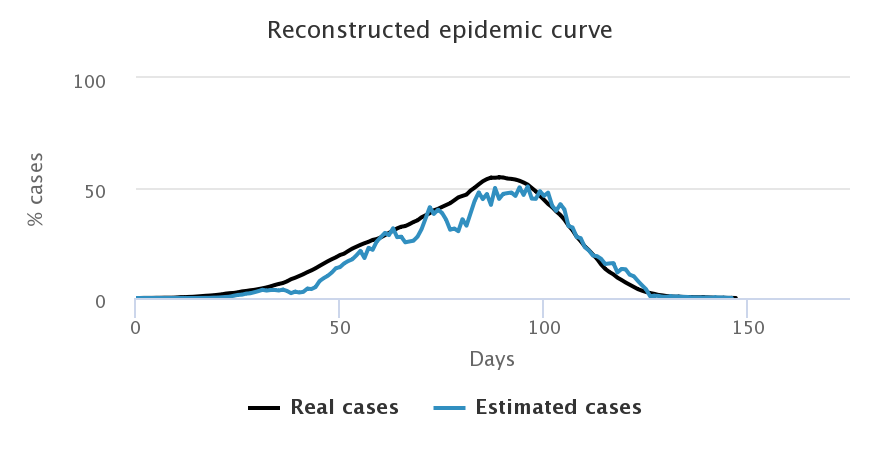}
        \centering \textbf{(b) 6 tests / 2000 people}
    \end{minipage}
    \begin{minipage}[b]{0.3\linewidth}
        \includegraphics[scale=0.22]{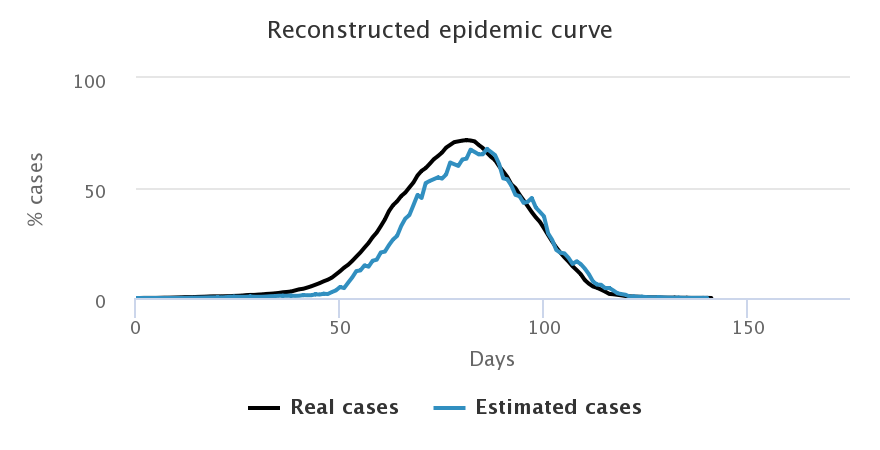}
        \centering \textbf{(c) 9 tests / 2000 people}
    \end{minipage}

    \caption{Experiment 2: immediate start, random sample, varying number of daily tests}
    \label{fig:exp2}
\end{figure*}

In a second experiment, we set the strategy to random sampling, and the beginning of the testing campaign as soon as the first case appears, and vary the \textbf{number of daily tests} from 3 tests per 2000 people (left), to 6 tests per 2000 people (middle) and finally 9 tests per 2000 people (right). Figure~\ref{fig:exp2} illustrates the resulting estimations of the epidemics curve, only with the predictive values method since it is better than the proportionality method.


We notice that the higher the number of tests, the better the estimation of the number of cases in the general population. After an initial overestimation, when there are very few cases in the population and very little tests performed, the reconstructed curve follows rather precisely the actual epidemic curve, in particular around the peaking time, the key moment of the epidemic.

\subsection{Comparison of estimated cases \wrt activation date of testing campaign}

\begin{figure*}[hbt]
    \centering
    
    \begin{minipage}[b]{0.45\linewidth}
        \includegraphics[scale=0.23]{imgs/Q17-estim-random-3-0-en.png}
        \centering \textbf{(a) Immediate start but low intensity \\(3 tests/2000)}
    \end{minipage}
    \hfill
    \begin{minipage}[b]{0.45\linewidth}
        \includegraphics[scale=0.23]{imgs/Q17-estim-random-9-0-en.png}
        \centering \textbf{(b) Immediate start with high intensity \\(9 tests/2000)}
    \end{minipage}
    
    \begin{minipage}[b]{0.45\linewidth}
        \includegraphics[scale=0.23]{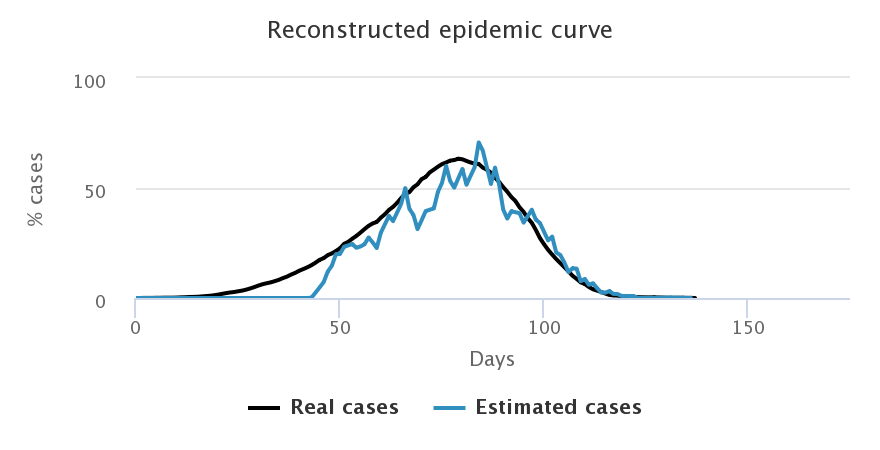}
        \centering \textbf{(c) Late start with low intensity \\(3 tests/2000)}
    \end{minipage}
    \hfill
    \begin{minipage}[b]{0.45\linewidth}
        \includegraphics[scale=0.23]{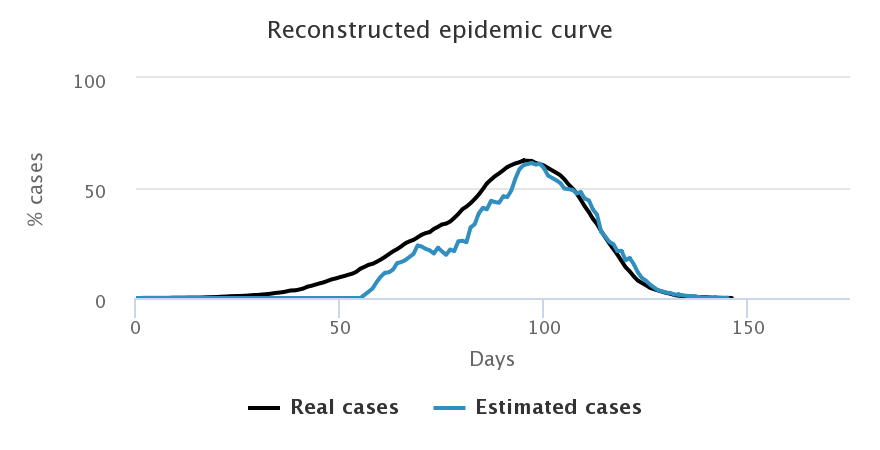}
        \centering \textbf{(d) Late start with high intensity \\(9 tests/2000)}
    \end{minipage}
    
    \caption{Experiment 3: random sample, varying starting date and intensity of the testing campaign}
    \label{fig:exp3}
\end{figure*}

In this final experiment, we use the random sample again, and vary the starting date of the campaign as well as its intensity, in terms of the number of daily tests performed. The lower intensity, 3 tests/2000 people, corresponds to the initial strategy in France, when the number of available kits was still quite low. The objective of this experiment is to assess how well we can estimate the \emph{''real''} epidemic curve, by either starting screening very early, or performing it at a very high intensity, or both. Figure~\ref{fig:exp3} shows the results of the 4 different combinations of parameters.

We notice that with a low screening intensity of 3 tests for 2,000 people (first column), the reconstructed curve cannot capture the epidemic peak when screening started late (after 15\% of infections, upper left). 

Similarly, if the screening campaign starts late (15\% people symptomatic, bottom figures), even increasing the number of tests to 9 per 2000 people (bottom right), we notice that the peak identification is very uncertain, and cases are strongly overestimated after the peak. When screening starts immediately and intensively, this overestimation only happens when the number of cases is very low (top right). The conclusion is that if we wait too late to activate the testing campaign, the knowledge of the epidemic is strongly reduced, even if we ramp up the number of daily tests.

\section{Model of vaccination}
In this section we describe the updated model including vaccination.

\subsection{Modifications to the population model}

The initial model was adapted to consider the new situation when the vaccine became available. In particular we lifted the lockdown rules that are not enforced anymore: concretely, individuals are not considered to work from home anymore, but instead their mobility is now affected by other factors.
Namely, all agents have 2 more attributes, randomly initialised based on their age:
\begin{itemize}
    \item Their \textbf{risk factor}, which represents their probability to develop a serious form of the illness when contaminated. Since age is a known comorbidity for COVID-19 mortality, the risk factor attribute is set randomly as proportional with age: most younger people have less risks to develop a serious form, even though some of them can have a high risk; and most older people have more risk to develop serious forms.
    \item Their number of \textbf{daily contacts}, that affects the number of other agents that can contaminate them at each step. This number is initialised with a random individual part, plus a random part reversely proportional to age (as contacts usually decrease with age), and minus another random part proportional to the level of risk (as we assume people who know they are at risk will voluntarily limit their contacts from fear of being ill).
\end{itemize}

We also added new states to the epidemic model, set as boolean variables of the agents:
\begin{itemize}
    \item \textbf{serious form}: when an agent is symptomatic it has a probability equal to its risk factor (or reduced if vaccinated) to develop a serious form. Serious forms last longer than other symptomatic forms. 
    %
    \item \textbf{vaccinated}: vaccinated agents have a lower probability (set by the vaccine efficiency parameter) to get infected, to contaminate others, and to develop a serious form when ill. 
\end{itemize} 


Finally, unlike the previous model, we now consider that the immunity conferred by either vaccination or recovery is not permanent. We added 2 parameters for these 2 durations of immunity. Agents then return to the Susceptible state, and their vaccinated attribute becomes false, meaning they can be infected again, and they can be considered for vaccination again. However, we remember for each agent the number of times they got sick, and the number of vaccine doses they received, for statistical purposes.

\subsection{Modelling vaccination}

The vaccination campaign has the following parameters: 
\begin{itemize}
    \item \textbf{Start} of the vaccination campaign, in percentage of infected people in the population.
    \item \textbf{Speed} of the vaccination campaign, in number of daily doses available.
    \item \textbf{Target} population that receives vaccine in priority, ("high risk", "more contacts", or "random").
    \item Vaccine \textbf{efficiency} was set to 90\%, meaning that vaccinated agents have a probability reduced by 90\%: to be infected when in contact with a sick person; to develop a serious form when infected and symptomatic; and to transmit the virus to their contacts when infected. To simplify we consider that the vaccine efficiency is the same on all 3 values.
    \item Duration of \textbf{immunity} conferred by the vaccine, after which the agent's status is reset to unvaccinated. 
\end{itemize}

If the proportion of sick agents in the population is above the selected threshold, the vaccination campaign can start. The target population for vaccination at each step is computed as follows. First, only agents that are not infected, and not currently immune (whether from vaccination or recovery) are considered as candidates. Then, the number of daily doses (parameter) provides the maximal number of agents that can be vaccinated at that step. Finally, this number of agents are selected among the candidates, in decreasing order of the value of the attribute corresponding to the selected strategy (daily contacts or risk factor), or randomly if so selected. This ensures that when the agents with the highest priority are already vaccinated, the following agents in order of priority can also access vaccination.


\subsection{Vaccination strategies}


We defined three vaccination strategies in our model:
\begin{itemize}
    \item \textbf{Randomly} targeting non-vaccinated, non-sick agents. 
    \item \textbf{Higher risk} first: targeting people with the highest level of risk first, and then in decreasing order of risk. This is the most used strategy where elderly people or people with comorbidities are vaccinated first.
    \item \textbf{Higher contacts} first: targeting people with the highest number of daily contacts first, and then in decreasing order of daily contacts. This is partly applied when vaccinating medical staff first since they are exposed to many sick or vulnerable people. Our literature review discussed several simulation showing the efficiency of this strategy.
\end{itemize}

\subsection{Interactive simulator}

\begin{figure}
    \centering
    \includegraphics[scale=0.23]{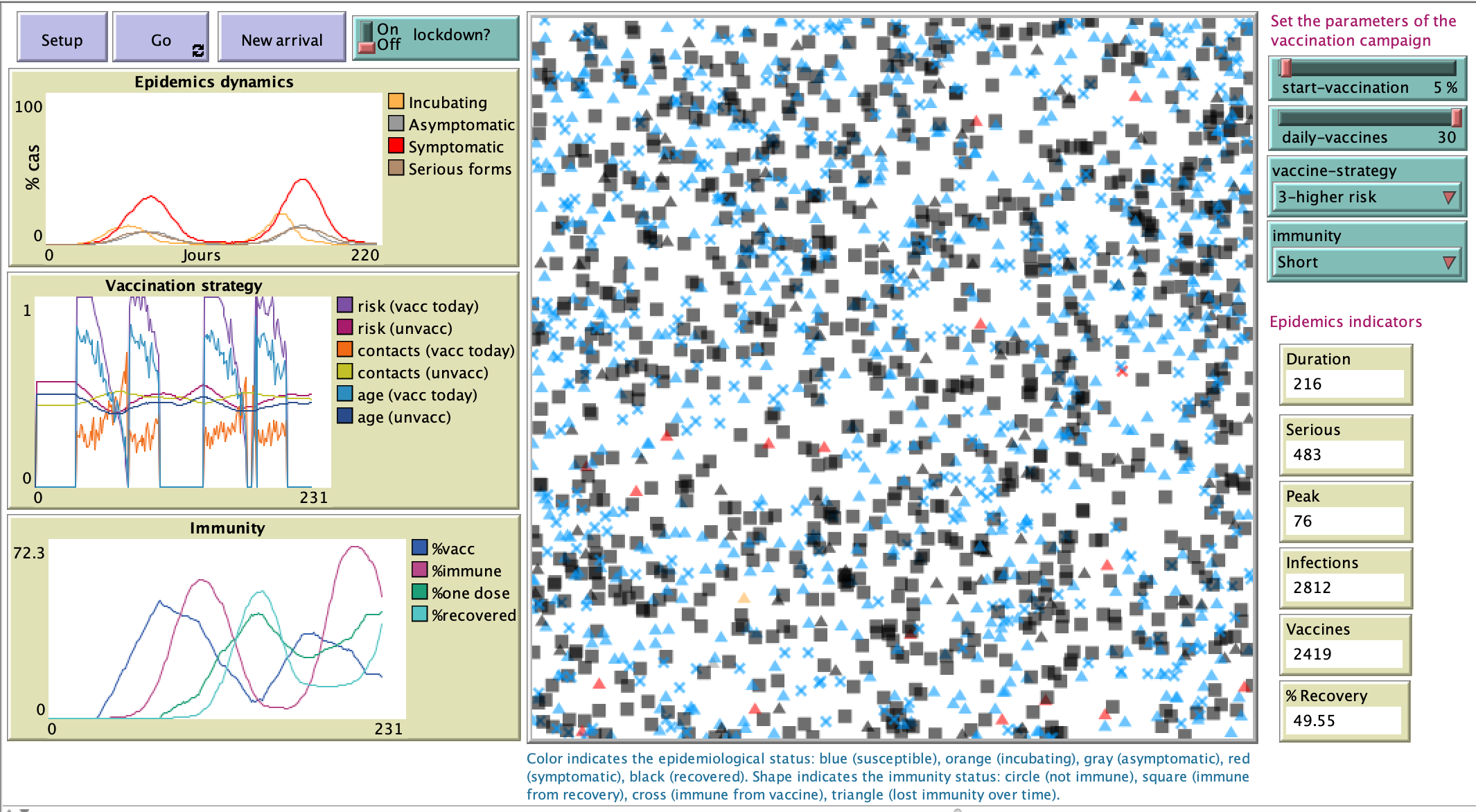}
    \caption{Screenshot of the online vaccination simulator}
    \label{fig:screen2}
\end{figure}

We implemented this new model in Netlogo\footnote{The model code and simulator are available online at: \url{http://nausikaa.net/wp-content/uploads/2022/10/virus2-vaccine-web-en.html}.}. The interface is rather similar to the previous simulator, but we have replaced the testing parameters with vaccination parameters. Our goal was to keep the interface simple enough, and not lose the user with too many different parameters. The fewer parameters there are, the easier it is to measure their individual impact. The user can select when to start the vaccination campaign, how many vaccines are available per day, the duration of immunity, and the target population. A number of graphs and monitors let them observe the results of their choices. Figure~\ref{fig:screen2} shows this interface.
We have also added 2 buttons: 
\begin{itemize}
    \item One button allows to randomly introduce a sick individual. This allows to simulate people arriving from the outside, who can possibly bring back the virus when the epidemics is locally controlled.
    \item One button allows to start/stop a lockdown: all agents stop moving until the lockdown is lifted. This allows to simulate emergency measures when the situation gets out of control.
\end{itemize}

\section{Vaccination experiments}

We conducted a number of experiments: first to test our model, and then to compare various parameters of the vaccination strategy.

\subsection{Verifying the impact of attributes}

\begin{figure}[hbt]
    \centering
\includegraphics[scale=0.35]{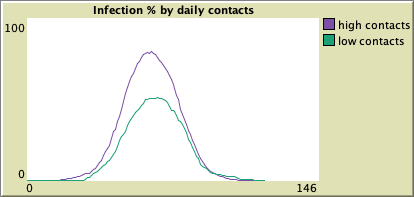}
\includegraphics[scale=0.35]{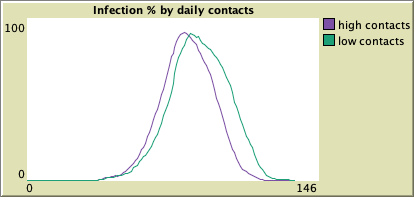}
    \caption{Infections by number of daily contacts: low incidence (top) or high incidence (bottom)}
    \label{fig:continf}
\end{figure}

\paragraph{Daily contacts.} We first tracked infections by number of daily contacts. The "low contacts" category are agents with less than 4 daily contacts, while "high contacts" are agents with over 10 daily contacts. We tracked the percentage of agents in each category who got infected. Figure \ref{fig:continf} shows that people with more contacts get infected first, and earlier than people with less contacts; but after a while, when the incidence rate is higher, even people with few contacts get infected as the virus has propagated to the whole population of agents. When vaccination is on, the incidence rate is lower, so the difference is even more visible between agents with high or low number of contacts.

\begin{figure}[hbt]
    \centering
\includegraphics[scale=0.4]{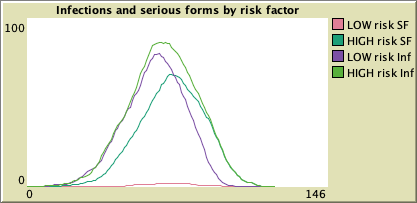}
    \caption{Infections and serious forms per risk category}
    \label{fig:riskinf}
\end{figure}

\paragraph{Risk factor.} We then tracked infections and serious forms by individual risk factor. We considered as "high risk" agents with a risk over 0.75, and as "low risk" agents with a risk under 0.25. We then tracked the percentage of agents in each category who got infected, and who developed a serious form of the illness. Figure~\ref{fig:riskinf} reports the results. We can see that there are slightly more agents infected among the "high risk" category, but the main difference is in the proportion of serious forms: almost none among "low risk" agents, but many among "high risk" agents. This also reflects in the width of the peak: since serious forms last longer than other infections, the peak is larger (lasts more steps) among "high" risk" than "low risk" agents.


\subsection{Verifying the vaccinations targets}

\begin{figure}
    \centering
\includegraphics[scale=0.55]{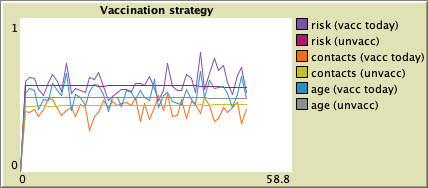}
\includegraphics[scale=0.55]{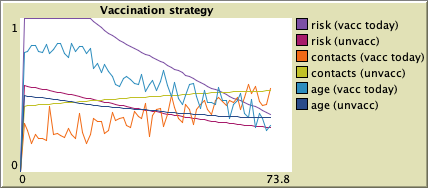}
\includegraphics[scale=0.55]{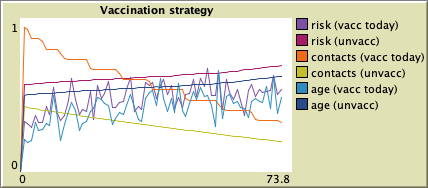}
    \caption{Vaccination targeting: random (top), risk-first (middle) and contacts-first (bottom).}
    \label{fig:vacc3targets}
\end{figure}

Figures~\ref{fig:vacc3targets} illustrate how the mean values of relevant attributes (daily contacts, risk factor, and age) evolve over the vaccination campaign. We averaged those three values in the group of agents that were the most recently vaccinated (at the previous step) as well as among the agents not yet vaccinated. 

\paragraph{Risk first.} One can see that with risk-first strategy (middle), the average risk factor of just vaccinated agents is stable at 1 (the maximum value) for an amount of time, requested to vaccinate all higher-risk agents. It then decreases as agents selected to receive vaccination have a lower and lower risk. In the mean time the level of risk of agents not vaccinated yet (and therefore still exposed to the virus) decreases. The average age also goes down as it is linked with a higher risk.

\paragraph{Contacts first.} When selecting agents with more contacts to be vaccinated first (bottom), one can observe that the average number of daily contacts decreases among agents just vaccinated, and not yet vaccinated. This means that the agents having more contacts, and more probability to get infected and/or to infect others, are already protected by the vaccine. We can observe that in that case the average age of vaccinated agents increases with time: since a younger age is linked to more contacts, younger people are vaccinated first.

\paragraph{Randomly.} The baseline is provided when vaccinating agents in random order (top). In that case we can see that the average values of attributes is oscillating around the mean value in the global population.

\subsection{Comparing durations of immunity}

\begin{figure}
    \centering
    \includegraphics[scale=0.5]{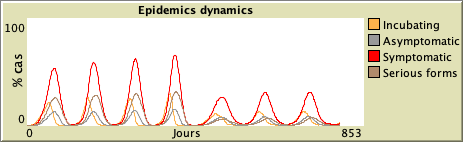}
    \includegraphics[scale=0.6]{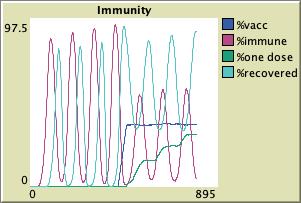}
    \caption{Epidemics dynamics and immunity of agents: 7 waves created with a shorter immunity duration (40 steps), vaccinating risk-first}
    \label{fig:7waves}
\end{figure}

In the following experiment, we set the duration of immunity provided by the vaccine, or by recovering from the illness, to 40 steps. We started without vaccination, and observed 4 successive waves of the epidemics. We then started vaccinating with the risk-first strategy, and observed 3 subsequent waves, but much lower than the previous ones. We manually stopped the simulation, since the stop condition that no agent is infected anymore cannot be reached with such a short immunity duration. Figure~\ref{fig:7waves} shows the epidemics dynamics with its 7 waves, as well as the immunity of the agents: agents that are protected after recovering, or after getting the vaccine, and agents that are susceptible again, despite having at least one dose of the vaccine, or one recovery.

This immunity duration of 40 steps is very short, but in a closed environment with no agents arriving from the outside, it is required to observe such waves. Another option to create such waves with a longer immunity duration is to introduce new infected agents in the simulation when the population of agents has become susceptible again. This is what happens in real life, when people start travelling again: even when the epidemics is locally controlled, new individuals can arrive and bring back the virus, which restarts another epidemics wave. We have witnessed this same phenomenon over and over again since the end of the first general lockdown. We have therefore added a button in the interface that lets the user add new infected agents whenever wanted. We could also extend the model to automatically randomly introduce such visitors from time to time in the simulation loop.




\subsection{Exhaustively comparing vaccination strategies}


\paragraph{Experimental setting} 
In order to exhaustively compare strategies, we ran various experiments, varying different parameters of the vaccination campaign. We have set the vaccine efficiency to 0.9 and immunity duration to permanent (to avoid infinite waves and ensure simulations stop). The simulation stops when no more agent is infectious. For each configuration, we ran 20 simulations in order to smooth randomness, and logged the average values of 5 indicators that allow to evaluate the efficiency of the campaign:
\begin{itemize}
    \item \textbf{Duration} of the simulation, \ie how long it took to control the epidemics
    \item \textbf{Serious}: number of serious forms, \ie how many agents got seriously ill
    \item \textbf{Peak}: epidemic peak as the maximal number of daily new infections
    \item \textbf{Infected}: total number of agents infected
    \item \textbf{Vaccines}: total number of vaccines injected
\end{itemize}
The results are summarised in the tables below.

\paragraph{Varying the strategy}
In this first experiment, we set the vaccination to start at a threshold of 5\% infected agents, with 30 daily vaccines. We then varied the strategy among the 3 possible targets. Table~\ref{tab:vary-strategy} shows the resulting indicators averaged over 20 runs per strategy.

\begin{table}[hbt]
\hspace*{-15pt}    \centering
    \begin{tabular}{|c|c|c|c|c|c|c|}
    \hline
    Strategy & Duration & Serious & Peak  & Infected & Vaccines \\
	    \hline
    Random   & 122,8    & 333,25  & 47,65 & 1156,4   & 936,9    \\
        \hline    
    Contacts & 132,15   & 401,55  & 52,55 & 1231,05  & 1011,45  \\
        \hline
    Risk     & 122,3    & 192,45  & 50,4 & 1163,55   & 914,3    \\
	    \hline
    \end{tabular}
    \caption{Comparing 3 strategies in terms of 5 indicators}
    \label{tab:vary-strategy}
\end{table}

\paragraph{Vaccinating high risk first}
In the next experiment, we set the strategy and varied its timing, \ie the start and intensity (number of daily doses) of the vaccination campaign. Table~\ref{tab:risk-stats} shows the results for the "risk first" strategy. We can observe that vaccinating the most vulnerable people very early (before any agent is infected, so a preventive campaign) is very effective against serious forms. With the start at time 0 and 30 daily doses, the number of serious cases is as low as 19.7 in average over 2000 agents. This is due to the much lower number of infections.

When starting later, the number of serious forms is much higher since many people would have been infected before the vaccination starts. This can be seen through the higher values of the total number of infections, and the subsequently lower number of vaccines injected (since recovered agents are already immune and not considered for vaccination).

\begin{table*}[hbt]
    \centering
    \begin{tabular}{|c|c|c|c|c|c|c|}
    \hline
    Start & Speed & Duration & Serious & Peak & Infected & Vaccines \\
    \hline
    \hline
    0     & 10    & 132,55   & 232,55 & 56,3  & 1501,55 & 695,6   \\
    \hline
    0     & 20    & 135      & 60,75  & 30,3  & 755,2   & 1408,15  \\
    \hline
    0     &	30    & 125,3    & 19,7   & 14,35 & 315,85  & 1769     \\
    \hline
    \hline
    10    & 10    & 127,2    & 472,35 & 75,8  & 1808,85 & 308,5    \\
    \hline
    10    & 20    & 124,35   & 339,45 & 71    & 1594,85 & 581,55   \\
    \hline
    10    & 30    & 126,6    & 275,45 & 61,15 & 1398,25 & 793,25   \\

    \hline
    \hline
    
    20    & 10    & 128,45   & 513,5  & 74,25 & 1828,6 & 274,25    \\
    \hline
    20    & 20    & 125,05   & 414,45 & 75,85 & 1674,4 & 473,4     \\
    \hline
    20    & 30    & 124,2    & 344,35 & 71,65 & 1530,25 & 644,7    \\
    
    \hline
    \hline
    \end{tabular}
    \caption{Risk-first vaccination, impact of early/late start and speed of vaccination on epidemic peak and serious forms}
    \label{tab:risk-stats}
\end{table*}

\paragraph{Contacts first.}
The results for the "more contacts first" strategy show that the number of serious forms is much higher with that strategy. It also shows again that when starting vaccination very early and fast, the epidemics can be contained to a very low level (with an epidemic peak of only 8.8 new daily infections). The difference in total number of infections is flagrant when starting immediately and increasing the speed: this shows how fast the epidemic can spread, so that going from 10 to 30 daily doses (for a population of 2000) makes an enormous difference. This difference is less flagrant when starting too late, since agents are already infected, as shown by the much lower number of vaccine doses used.
\begin{table*}[hbt]
    \centering
    \begin{tabular}{|c|c|c|c|c|c|c|}
    \hline
    Start & Speed & Duration & Serious & Peak & Infected & Vaccines \\
    \hline
    \hline
0  & 10 & 143,9  & 500,4  & 51,5  & 1502,35 & 778,1    \\
    \hline
0  & 20 & 145,6  & 167,1  & 17,7  & 505,5   & 1627,6   \\
    \hline
0  & 30 & 122    & 57     & 8,8   & 176,8   & 1877,25  \\
 \hline
    \hline
10 & 10 & 134,8  & 619    & 75    & 1837,5  & 339,95   \\
    \hline
10 & 20 & 133,45 & 519,4  & 67,85 & 1590    & 637,65   \\
    \hline
10 & 30 & 128,15 & 443,65 & 58,25 & 1362,35 & 867,95   \\
 \hline
    \hline
20 & 10 & 131,55 & 633,55 & 78,65 & 1874,25 & 275,55   \\
    \hline
20 & 20 & 128,75 & 560,85 & 71,4  & 1687,15 & 512,45   \\
    \hline
20 & 30 & 128,15 & 470,25 & 65,4  & 1463,3  & 646,7    \\
 \hline
    \hline
    \end{tabular}
    \caption{Contacts-first vaccination, impact of early/late start and speed of vaccination on epidemic peak and serious forms}
    \label{tab:contacts-stats}
\end{table*}

\section{Discussion and conclusion} \label{sec:cci}

\subsection{Summary of contributions}
In this work, we have provided a simple model of the propagation of COVID-19 in a population. In a first version, we have integrated different screening strategies to both control and evaluate the epidemic curve. The model is based on figures available for France but is easily adaptable to other countries. It is willingly simplified, in order to provide an interactive simulator to explain to the general public the mechanisms of the epidemics and how the count of infected people is estimated every day. Besides, we have also used our simulator to compare different computation methods to estimate the epidemic curve, and different screening parameters. In other work \cite{iscram2022}, we have used an optimisation algorithm to prove the required properties of a successful screening campaign, which happen to have been successfully used by South Korea at the start of the epidemics.

In a second version of our model, we have integrated several vaccination strategies to either limit the total number of cases by targeting people with more contacts, or limit the number of serious forms by targeting people with risk factors. Consistently with other works, we find that vaccinating younger people very early on can provide a better herd immunity to limit the total number of cases. However, we also show that when starting vaccination later or when the immunity duration is shorter, it is important to rather vaccinate older people who have more risks factors, in order to reduce the number of serious forms.

Both simulators are available online, in French and English, so that people can play with them. The source code is also available.
We believe that providing the population with scientific facts and explanations is key to protect them from fake news and to improve the acceptability of sanitary measures (\eg necessity of regular self-tests for children at school, usefulness of vaccination, etc). 

\subsection{Limitations and future work}
In the interest of simplicity of use, we have limited the number of parameters and outputs in each simulator. As a result, we have not tested the combination of sanitary measures (screening + quarantine, vaccination). Another limitation concerns trust: we assumed here that agents eligible to vaccination will receive the vaccine in order of priority. In fact, this is not necessarily true, some people might not want to receive the vaccine because they do not trust it. In other work, we are interested in how trust evolves and influences the speed of the vaccination campaign \cite{gretsi2022}.

\enlargethispage{20pt}


Our simulator has been made available online but we have not yet surveyed the users to assess its impact. Future experiments need to be setup with users of different profiles to prove our claim that changing role (playing the role of a health decider), actually interacting with the parameters to test what-if scenarios, and getting feedback about one's choices, can provide a welcome feeling of control and a better understanding and acceptance of a difficult situation.

\begin{acks}
We would like to thank to the CovPrehension collective\footnote{\url{https://covprehension.net}} for their insight and help about design and testing of the screening model, as well as publication and translation of the original blog post.

\end{acks}


\footnotesize


\end{document}